\documentclass[aps,prl,showpacs,preprint,floatfix]{revtex4}\usepackage{latexsym}
\usepackage{latexsym}
\usepackage{amsfonts}
\usepackage{amssymb}
\usepackage{amsmath}
\usepackage{graphicx}

\begin{document}

\title{Origin of the spectral linewidth in non linear spin-transfer oscillators based on MgO tunnel junctions}

\author{B. Georges, J. Grollier, V. Cros, A. Fert}
\affiliation{Unit\'e Mixte de Physique CNRS/Thales and Universit\'e
Paris Sud 11, Route
d\'epartementale 128, 91767 Palaiseau, France}
\author{A. Fukushima, H. Kubota, K. Yakushijin, S. Yuasa, K. Ando}
\affiliation{National Institute of Advanced Industrial Science and Technology (AIST) 1-1-1 Umezono, Tsukuba, Ibaraki 305-8568, Japan}

\begin{abstract}
We demonstrate the strong impact of the oscillator non-linearity on the line broadening by studying spin transfer induced microwave emission in MgO-based tunnel junctions as a function of both the injected dc current and the temperature. In addition, we give clear evidences that the intrinsic noise is not dominated by thermal fluctuations but rather by the chaotization of the magnetic system induced by the spin transfer torque. A consequence is that the spectral linewidth is almost not reduced in decreasing the temperature. 
\end{abstract}

\maketitle
The microwave emission associated with spin transfer induced magnetization precessions in metallic magnetic nanostructures leads to very promising possibilities for the development of new nanoscale microwave oscillators. Many experimental and theoretical studies have been initiated (see Stiles and Miltat \cite{Stiles} and references therein) to improve the sample characteristics in order to optimize the microwave properties of these nano-devices, in particular in terms of output power. In this vein, the recent development of low resistance MgO barriers \cite{Parkin, Yuasa} has allowed the injection of the necessary high current densities to manipulate the magnetization through the spin transfer effect \cite{Slonc, Berger} in magnetic tunnel junctions (MTJs). Sustained oscillations of the magnetization in such MTJs are of great interest since the power scales with the magneto-resistance ratio (MR) that is typically 100$\%$ in these devices at room temperature. For standard excitations in the free magnetic layer, output powers up to 1 $\mu W$ have been measured for a single spin transfer nano-oscillator (STNO)\cite{Deac, Petit, Nazarov, Houssameddine, Devolder, Cornelissen}. Further improvements of the output power will probably go through the synchronization of many of these oscillators \cite{GeorgesAPL}. However this objective might be questioned because of the observed peak linewidths (larger than 100 MHz) that are detrimental to reach a phase locked state  \cite{GeorgesPRL}. To go beyond this strong drawback, a fundamental study has to be led to determine the mechanisms at the origin of the peak linewidth in MTJs.

In this Letter, we present an experimental study of the microwave emission in MgO based MTJs. From the dependence with the dc current, we show the strong impact of the high non-linearity of the oscillator on the linewidth, as predicted by the recent theory of STNOs \cite{SlavinIEEE, TiberAPL, KimPRL08, KimPRL, Tiber}. Line broadening is also related to the different sources of noise. From the temperature ($T$) dependence of the linewidth, we evidence that the dissipation process is not dominated by thermal fluctuations but rather by a spin transfer induced noise.

Our magnetic tunnel junctions are composed of PtMn 15/ CoFe 2.5 / Ru 0.85 / CoFeB 3 / MgO 1.075 / CoFeB 2 (nm) and patterned into an elliptical shape of dimension 170x70 $nm^{2}$ \cite{Nagamine}. The RA product is 0.85 $\Omega$.$\mu$$m^{2}$ for the parallel (P) magnetization configuration at $T$ = 300 K. The tunnel magneto-resistance ratio (TMR) is 100$\%$ at 300 K and 140$\%$ at 20 K. The results are obtained with a magnetic field $H$ between 100 and 300 Oe, applied along the easy axis of the ellipse that stabilizes the antiparallel (AP) configuration. The switching field of the free magnetization from P to AP (AP to P) occurs at 38 Oe (-25 Oe). The junctions are biased with a dc current ($I_{dc}$) ranging from 0.3 to 1.8 mA that destabilizes the AP configuration. Even for the largest value of $I_{dc}$, no modification of the barrier quality is observed. Microwave measurements up to 10 GHz are recorded on a spectrum analyzer after 35 dB amplification. The background noise, obtained at $I_{dc}$ = 0, is substracted to the power spectra.

\begin{figure}[h]
   \centering
    \includegraphics[width=8.5 cm]{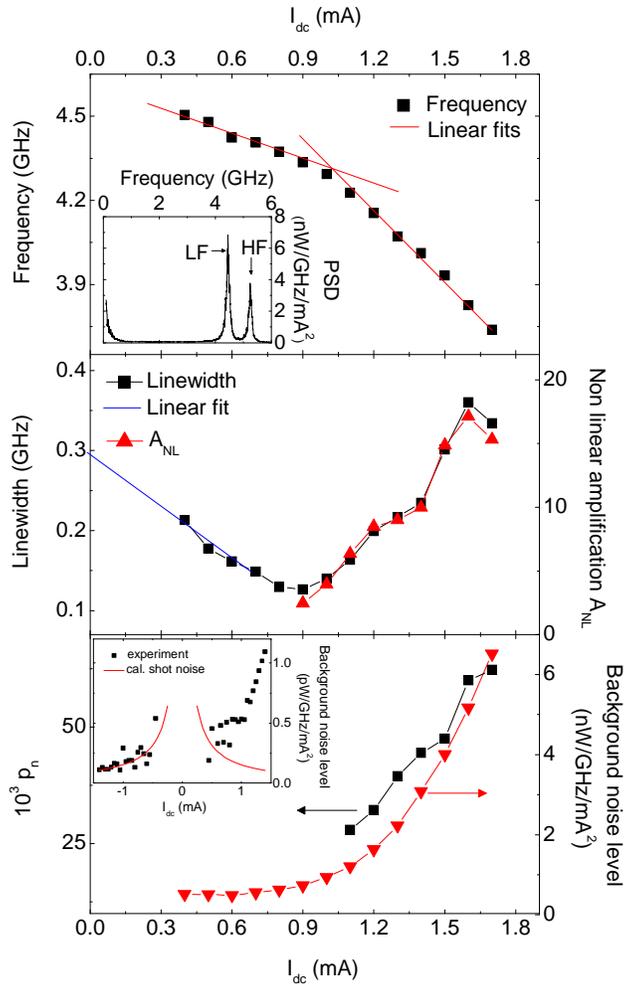}
     \caption{Georges \textit{et al.}}
\label{fig1}
\end{figure}

The power spectra are characterized by two well-separated peaks, labeled low frequency (LF) and high frequency (HF) modes together with a large 1/f noise (see inset of Fig.\ref{fig1}(a)). As mentioned in Ref.\cite{Deac}, LF and HF modes correspond respectively to a center and edge modes of the ellipse. In this Letter, microwave features of the LF are shown. Similar behaviors are obtained for the HF mode. In Fig.\ref{fig1}(a), we display the change of the frequency $f_0$ of the LF mode with $I_{dc}$ for $H$ = 110 Oe at $T$ = 300 K. The overall frequency red shift is characteristic of an in-plane oscillation of the magnetization \cite{Krivorotov}. In Fig.\ref{fig1}(b) we show the corresponding variation of the peak linewidth with $I_{dc}$, that depicts two different regimes. Below a threshold current $I_{th}$ $\approx$ 1 mA, the linewidth decreases with $I_{dc}$ while above that value it increases strongly. First, we focus on the low current regime ($I_{dc} < I_{th}$) in which the frequency decreases slowly (see Fig.\ref{fig1}(a)). In the recent theoretical description of STNOs \cite{KimPRL}, this regime is associated with thermally excited ferromagnetic resonance (FMR) noise for which no variation of the frequency is expected. Our experimental decrease of $f_0$ can be attributed to the current dependent torques due to the Oersted field and/or field-like torque \cite{KubotaSankeyFuchs}. In this regime, a strong reduction of the linewidth down to a minimum of 120 MHz at 0.9 mA is measured as shown in Fig.\ref{fig1}(b). This behavior is related to the gradual compensation of the natural damping of the magnetization by the spin transfer torque. For a classical STNO \cite{Tiber}, a linear decrease of the linewidth with $I_{dc}$ is expected:  $\Delta$f = $\Gamma_{g} - \frac{\sigma}{2\pi}$$I_{dc}$, where $\Gamma_{g} \approx \alpha$$\frac{\gamma \mu_{0} M_{eff}}{2\pi}$ represents the natural FMR linewidth in the case of an in-plane magnetic field \cite{KimPRL}, $\alpha$ is the Gilbert damping, $\gamma$ is the gyromagnetic constant, $\mu_{0} M_{eff}$ is the effective magnetization and $\sigma$ is related to the spin transfer efficiency \cite{SlavinIEEE}. From a linear extrapolation at zero current of the linewidth (see blue fitting line in Fig.\ref{fig1}(b)), we obtain $\Gamma_{g}$ = 0.3 GHz. From the frequency dependence on the magnetic field (not shwon) that follows the Kittel formula, we estimate $\mu_{0} M_{eff}$ = 1.16 T. We then deduce the effective damping parameter $\alpha$ = 0.009 $\pm$ 0.004. This value agrees with the measured damping parameter (0.013) of the 2 nm thick CoFeB layer, obtained by FMR experiments on the unpatterned junction stack. 

In the second regime ($I_{dc} > I_{th}$),  the steep increase of the linewidth with $I_{dc}$ (Fig.\ref{fig1} (b)) is associated with a stronger decrease of the frequency (Fig.\ref{fig1} (a)). This behavior is characteristic of non linear oscillations sustained by the spin transfer torque \cite{SlavinIEEE}. Assuming that the non linear damping term $Q$ is zero the linewidth is expressed as \cite{KimPRL08}:
\begin{equation}\label{LWlowT}
\Delta f = A_{NL}\times \Gamma_{g} \frac{P_{n}}{E(p_{0})}
\end{equation}
\begin{equation}\label{ANL}
A_{NL} = 1+\left(\frac{I_{dc}}{\Gamma_{g}}\frac{df}{dI_{dc}}\right)^{2}
\end{equation}
where $\frac{df}{dI_{dc}}$ is the agility in current, $P_{n}$ is the noise amplitude and $E(p_{0})$ is the oscillator energy.  The first term $A_{NL}$ describes the phase noise amplification due to the non linearity which is related to the oscillator agility in current. In Fig.~\ref{fig1}(b), we show , the variation of the calculated $A_{NL}$ with $I_{dc}$ in the above-threshold regime, using the experimental variation of $\frac{df}{dI_{dc}}$ and $\Gamma_{g}$. It reproduces very well the evolution of the linewidth with $I_{dc}$, thus confirming the strong impact of the non linearity on the peak broadening.

The second term $\Gamma_{g} \frac{P_{n}}{E(p_{0})}$, in Eq.(\ref{LWlowT}), is the normalized phase noise that corresponds to the generation linewidth of a $"linear"$ auto-oscillator, for which the fluctuation-dissipation theory predicts a constant noise level ($P_{n} = k_{B}T$) \cite{Kubo}. Furthermore, the oscillator energy $E(p_{0})$ is proportional to the emitted power $p_0$ \cite{TiberAPL}. We calculate $p_0$ as $p_{0} = \left[p_{int} - p_{int} (min)\right]/p_{int}(min)$, where $p_{int}$ is the peak integrated power normalized by $\left[\left(R_{AP}-R_{P}\right)/\left(R_{AP}+R_{P}\right)\right]^{2}\left(I_{dc}\right)^{2}$ to take into account the bias dependence of the resistances and the increase of the emitted power amplitude with $\left(I_{dc}\right)^{2}$ \cite{Deac}. Then we calculate from Eq.(\ref{LWlowT}) the variation of $p_{n} = \Delta f p_0 / A_{NL} \Gamma_g$ that is proportional to $P_{n}$ (see black squares in Fig.\ref{fig1}(c)). We observe a significant increase of the calculated noise level $p_{n}$ with $I_{dc}$ in contrast with the expected constant noise level $P_{n} = k_{B}T$. In Fig.\ref{fig1}(c), we compare these calculated values to the background level of the power spectra  taken between 2 and 3 GHz. This background noise measurements represent another way to probe the noise amplitude. This noise level increases similarly to $p_n$, confirming that the noise amplitude is not constant.

We display in the inset of Fig.\ref{fig1}(c) the measured background noise for both current polarities. The clear observed asymmetry in current allows us to discard some possible sources of noise in MTJs. The first one is the Joule heating that has actually a minor impact on the effective temperature. Indeed, in order to estimate the current induced heating in our device, we measure the switching field (at about -1000 Oe) of the synthetic antiferromagnet at $I_{dc}$ = 0.1 mA as a function of the temperature (not shown). This switching field decreases linearly with the temperature at a rate of 1.2 Oe/K. Then we measure this switching field as a function of $I_{dc}$ at 20 K and estimate a temperature increase of about 25 K for $I_{dc}$ = 1.7 mA. Another source of current symmetric noise in MTJs is the shot noise. With our experimental conditions of applied voltage and temperature, it is expressed as $2eI\left(dV/dI\right)^{2}$ \cite{Klaassen}. We display the calculated shot noise (divided by $I_{dc}^{2}$) as a function of $I_{dc}$ (see inset Fig.\ref{fig1}(c)). We observe that at negative current, for which the spin transfer torque stabilizes the magnetization, the evolution of the background noise level is well reproduced by the calculated shot noise. On the contrary, at positive currents, the background noise level increases largely above the shot noise level. 

As the large increase of the background noise level occurs for $I_{dc} > I_{th}$, we believe that the spin transfer torque is responsible for such noise enhancement. Several types of spin dependent mechanism may occur in magneto-resistive devices. On the one hand, Chudnovskiy \textit{et al.} \cite{Chudnovskiy} have calculated that the spin torque shot noise, related to fluctuation of dc current polarization direction, may be important in MTJs. However this mechanism should be independent on the current polarity \cite{Foros}. On the other hand, spin torque dependent noise may also have its origin in the excitation of incoherent spin-waves \cite{ZhuIEEE}. In all metallic devices, such as GMR read heads, noise measurements have been performed only in the low frequency range (up to 100 MHz) \cite{Smith}. It is observed that the noise is also highly asymmetric in current. Smith \textit{et al.} predict that this mag-noise appears below the FMR peak frequency. In our devices, we measure this asymmetry for the 1/f noise but also for the background noise, well above the LF and HF peaks. An important issue is to understand whether this spin dependent noise is specific to MTJs (since smaller linewidths are measured in metallic devices \cite{Krivorotov}), or only related to complex dynamics of the magnetic system. In metallic devices, large dc current are injected, creating a stronger Oersted field that could explain the excitation of different modes compared to MTJs. Another characteristic of MTJs is the possible existence of hot spots in the insulating barrier that leads to spatially inhomogeneous current densities, thus enhancing the incoherence of the magnetic system. Finally magneto-resistance ratio in MTJs are much larger than in metallic systems. Therefore, significant spatial fluctuations of the current and/or its spin polarization can generate an additional magnetic noise through the spin transfer torque.

\begin{figure}[h]
   \centering
    \includegraphics[width=8.5 cm]{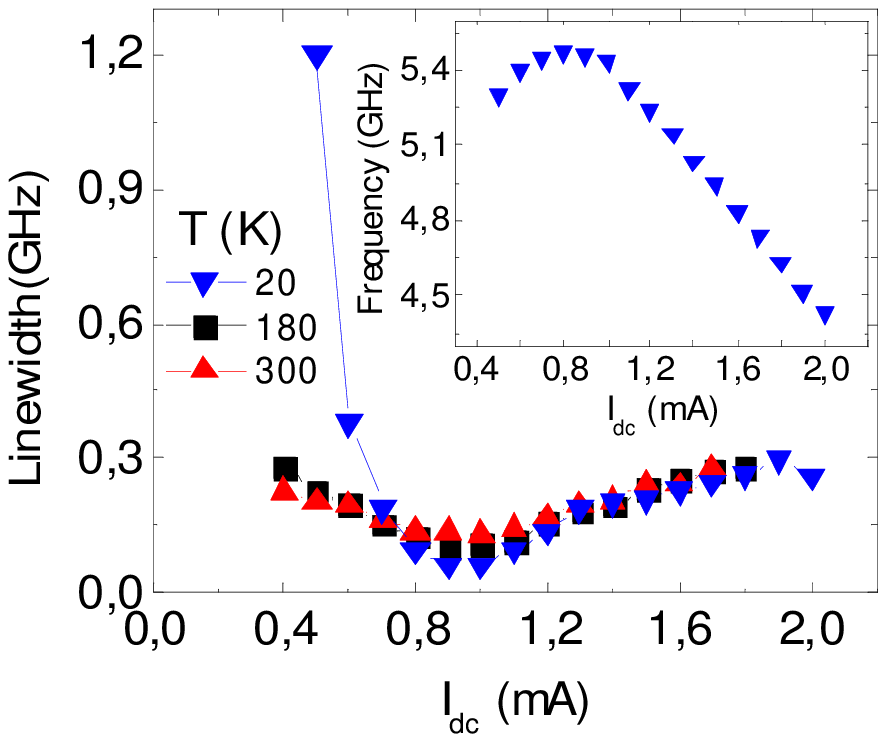}
     \caption{Georges \textit{et al.}}
\label{fig2}
\end{figure}

\begin{figure}[h]
   \centering
    \includegraphics[width=8.5 cm]{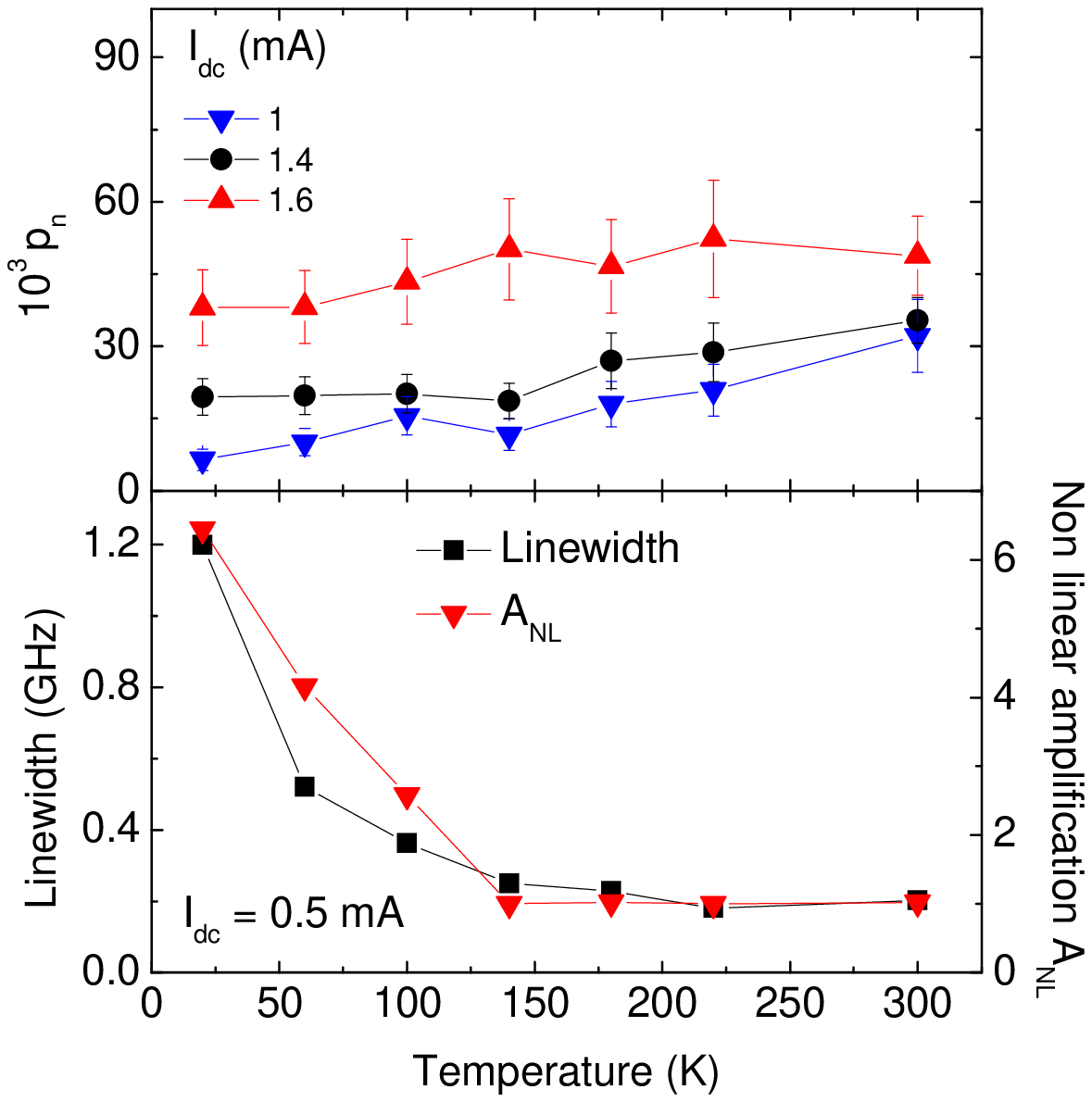}
     \caption{Georges \textit{et al.}}
\label{fig3}
\end{figure}

In order to investigate in more details this spin torque dependent noise, we have studied the microwave emission as a function of the temperature from 300 K down to 20 K. At all temperatures, the linewidth variation with $I_{dc}$ is characterized by the two regimes discussed previously (see Fig.\ref{fig2}). First, we focus on the above-threshold regime where the linewidth is almost unchanged with $T$. For each temperature, we calculate the noise level $p_{n}$ as described before. In Fig.\ref{fig3}(a), we show the resulting temperature dependence of $p_{n}$ for three current values above the threshold current: $I_{dc}$ = 1, 1.4 and 1.7 mA. The calculated noise level $p_{n}$ increases with $I_{dc}$ for all temperature. The observed weak increase of $p_{n}$ with $T$ for all currents discards that current fluctuations due to the large MR ratio are the dominant source of noise. Indeed, by this mechanism, the noise level $p_n$ should decrease with temperature as the magneto-resistance does, i.e. 15 $\%$ between 20 and 300 K at $I_{dc}$ = 1.7 mA. On the contrary, the weak increase of $p_n$ with $T$ could correspond to a higher magnetic stiffness at low temperature. However we can not rule out an impact of the noise originating from transport inhomogeneities due to hot spots that should be independent on temperature. 

In the below-threshold regime, the linewidth increases from 0.2 GHz to 1.2 GHz while decreasing the temperature from 300 to 20 K as observed in Fig.\ref{fig2} and specifically shown in Fig.\ref{fig3}(b) for $I_{dc}$ = 0.5 mA. This increase of the linewidth at low temperature and low currents goes along with a strong enhancement of the agility in current. The inset of Fig.\ref{fig2} shows the variation of the frequency with the dc current at $T$ = 20 K for $H$ = 205 Oe. In the below-threshold regime the frequency is strongly increasing with $I_{dc}$ whereas it is slowly decreasing at $T$ = 300 K. Then at low temperature there exists an additional unexpected agility in current that impacts the linewidth.  We show in Fig.\ref{fig3}(b) that, the non linear amplification parameter $A_{NL}$ calculated using Eq.\ref{ANL} behaves in temperature very similarly to the linewidth. To account for non linear effects, we propose to modify the standard expression of the linewidth in the below-threshold regime as follows: 

\begin{equation}\label{below}
\Delta f_{I_{dc} < I_{th}} = \left(\Gamma_{g} - \frac{\sigma}{2\pi} I_{dc} \right) \times A_{NL}
\end{equation} 

where the parameter $A_{NL}$ is the one used in the above-threshold regime given in Eq.\ref{ANL}. The mechanism at the origin of this agility is beyond the scope of this paper. However we can discard once again the effect of the Joule heating that would lead to a decrease of the effective magnetization and the frequency with $I_{dc}$. As this phenomena is current and temperature dependent, it might be related to some modifications of the transport mechanisms across the MgO barrier, being more coherent at low temperature, or to the field-like torque that can vary with the temperature \cite{Manchon}.

In conclusion, we have shown that the microwave emission induced by the spin transfer in MTJs are well described at a given temperature by the theory of non linear oscillators. The reduction of the linewidth in the below-threshold regime is characteristic of FMR-type excitations. At low temperature, the linewidth in this regime increases strongly. We describe this behavior in terms of an additional agility in current that amplifies the linewidth. In the above-threshold regime, the linewidth is strongly enhanced due to the non linear effect of the spin transfer induced precessions. Moreover, we demonstrate that spin torque dependent fluctuations are at the origin of the noise. By cooling down the system to 20 K, the linewidth is unexpectedly not decreasing significantly. Our analysis indicates that the excitations of incoherent magnetic modes and/or the presence of hot spots are probably at the origin of this unusual noise. 

The authors acknowledge H. Hurdequint for FMR measurements. B. G. is supported by a PhD grant from the DGA. This work is partially supported by the CNRS and ANR agency (NANOMASER PNANO-06-067-04).

\newpage

\textbf{Figure captions}\\

Figure 1. (Color Online) (a) \textit{Inset}: Representative power spectral density (PSD) normalized by $I_{dc}^{2}$, obtained for $I_{dc}$ = 1 mA and  $H$ = 110 Oe at $T$ = 300 K. Two large peaks are observed labeled low frequency (LF) and high frequency (HF) modes. \textit{Main panel}: variation of the frequency of the LF mode (black squares) with $I_{dc}$ for $H$ = 110 Oe at $T$ = 300 K . The lines are linear fits corresponding to the two regimes discussed in the text. (b) \textit{Left axis}: black squares represent the linewidth of the LF mode as a function of $I_{dc}$ for $H$ = 110 Oe and $T$ = 300 K. \textit{Right axis}: evolution of the calculated non linear amplification factor $A_{NL}$ (red triangles) with $I_{dc}$. (c) \textit{Left axis}:  dependence of the calculated $p_n = \Delta f p_0 / A_{NL} \Gamma_g$ (black squares) on $I_{dc}$. \textit{Right axis}: relative variation with $I_{dc}$ of the normalized background noise level (red triangles).\\

Figure 2. (Color Online) Variation of the linewidth with $I_{dc}$ for $T$ = 20, 180 and 300 K and $H$ = 205 Oe. \textit{Inset:} Variation of the frequency with $I_{dc}$ for $H$ = 205 Oe and $T$ = 20 K.\\

Figure 3. (Color Online) (a) Temperature dependence of the calculated noise level $p_n$ for $I_{dc}$ = 1, 1.4 and 1.7 mA for $H$ = 205 Oe. (b) \textit{Left axis}: Temperature variation of linewidth for $I_{dc}$ = 0.5 mA. \textit{Right axis}: Temperature variation of the non linear amplification parameter $A_{NL}$ calculated for $I_{dc}$ = 0.5 mA and $H$ = 205 Oe.\\

\newpage

\newpage

\newpage

\end{document}